\documentclass[preprint2]{aastex}
\newcommand\kms{km\,s$^{-1}$}
\newcommand\hash{$-$}
\begin{document}

\title{Large-Scale Velocity Structures in the
    Horologium-Reticulum Supercluster}

\author{Matthew C. Fleenor, James A. Rose, Wayne A. Christiansen}
\affil{Department of Physics \& Astronomy, University of North Carolina,
    Chapel Hill, NC 27599}
\email{fleenor2@physics.unc.edu, jim@physics.unc.edu, wayne@physics.unc.edu}

\author{Richard W. Hunstead}
\affil{School of Physics, University of Sydney, NSW 2006, Australia}
\email{rwh@physics.usyd.edu.au}

\author{Melanie Johnston-Hollitt}
\affil{Department of Physics, University of Tasmania, TAS 7005, Australia}
\email{Melanie.JohnstonHollitt@utas.edu.au}

\author{Michael J. Drinkwater}
\affil{Department of Physics, University of Queensland, QLD 4072, Australia}
\email{mjd@physics.uq.edu.au}

\and

\author{William Saunders}
\affil{Anglo-Australian Observatory, Epping NSW 1710, Australia }
\email{will@aaoepp.aao.gov.au}

\begin{abstract}
We present 547 optical redshifts obtained for galaxies in the region of the
Horologium-Reticulum Supercluster (HRS) using the 6dF multi-fiber 
spectrograph on the UK Schmidt Telescope at the Anglo Australian
Observatory. The HRS covers an area of more than $12\degr$ $\times$
$12\arcdeg$ on the sky centered at approximately $\alpha = 03^h19^m,
\delta = $ $-$50\degr 02\arcmin . Our 6dF observations concentrate
upon the inter-cluster regions of the HRS, from which we describe
four primary results. First, the HRS spans at least the redshift
range from 17,000 to 22,500 \kms.  Second, the overdensity of galaxies
in the inter-cluster regions of the HRS in this redshift range is
estimated to be 2.4, or $\delta \rho / \bar{\rho} \sim$ 1.4. Third, we find
a systematic trend  of increasing redshift along a Southeast-Northwest
(SE$-$NW) spatial axis in the HRS, in that the mean redshift of HRS
members increases by more than 1500 \kms\ from SE to NW over a
12\arcdeg\ region. Fourth, the HRS is bi-modal in redshift with a
separation of $\sim$2500 \kms\ (35 Mpc) between the higher and lower
redshift peaks. This fact is particularly evident if the above
spatial-redshift trend is fitted and removed.  In short, the HRS
appears to consist of two components in redshift space, each one
exhibiting a similar systematic spatial-redshift trend along a SE$-$NW
axis. Lastly, we compare these results from the HRS with the Shapley
supercluster and find similar properties and large-scale features. 

\end{abstract}

\keywords{galaxies: clusters: general, large-scale structure}

\section{Introduction}
Superclusters of galaxies represent the largest known conglomerations of both
visible and dark matter in the universe \citep{kal98}.  Given the complex
morphologies of superclusters
\citep[e.g.,][]{del86,hay86,wes95,bar98,bar00,dri04},  
as well as their huge scale \citep[e.g.,][]{zuc93,ein01} and potential
alignment within the local universe \citep{tul92}, superclusters pose
unique challenges for scenarios of the growth of and
inter-relationship between structures on all scales, such as the
hierarchical structure formation picture \citep{bau00,mot04}. Detailed
studies of the supercluster environment require extensive redshift
information over large areas of the sky, sampling both the intra- and
inter-cluster regions \citep{bar00}. Wide-field, multi-fiber
spectrographs are ideally suited to this task, as 
they permit three-dimensional probing of structures on megaparsec scales. 

The range of structures identified as superclusters varies widely in
terms of morphology and size. On the one hand there are superclusters
containing just a few major galaxy clusters connected by long
spiral-rich galaxy filaments. Two such examples are the Coma Supercluster 
\citep{gre78,del86,wes95} and the nearby
Pisces-Perseus Supercluster \citep{hay86,cha90}.  In contrast, other
structures are perhaps more readily characterized by the 
presence of rich clusters, from the few in the Hercules Supercluster
\citep{bar98} to those containing on the order of twenty major
clusters, as in the case of the Shapley supercluster 
\citep[e.g.,][]{bar00,qui00,dri99,dri04}.  In these latter cases, there is
evidently a rich variety of substructure present in these large-scale
entities.  Finally, \citet{tul92} have found that the superclusters
within the local universe exhibit a pronounced tendency to align
within the preferred plane of the Virgo Supercluster, representing a
structure on the scale of $\sim$0.1c. 

Originally noted by Shapley (1935) as exhibiting ``a considerable
departure from uniform distribution,'' the Horologium-Reticulum
Supercluster (HRS) is now recognized as one of the largest
superclusters in the local universe \citep{luc83, zuc93, ein01},
containing more 
than twenty Abell  clusters \citep[][hereafter ACO]{aco89}. The HRS 
covers an area of the sky in excess of 100 square degrees,
centered at approximately $\alpha = 03^h19^m, \delta = $
$-$50\degr02\arcmin\ \citep{zuc93}. In fact, in terms of mass
concentrations within 200 Mpc, the HRS stands as second
only to the Shapley supercluster \citep{hud99,ein01}.  It is of interest
to note that while the Shapley supercluster lies within the preferred
plane discussed by \citet{tul92}, the HRS lies more than 150 Mpc
outside of that plane.   

Recent studies in the HRS have focused exclusively upon the rich
clusters in the region. \citet{kat98} summarize the redshift
information from the ESO Nearby Abell Clusters Survey (ENACS), which
investigated ACO cluster cores throughout the HRS (specifically A3093,
A3108, A3111, A3112, A3128, A3144, and A3158). \citet{ros02} examined
the merging double-cluster system A3125/A3128, which is located in the
Southeast portion of the HRS. This multi-wavelength study revealed a
number of rapidly infalling groups and filaments, which were
accelerated by the HRS potential.  The results from their observations
imply that the HRS contains evolving substructures on a wide range of
mass scales.  

To date, no studies have been carried out that concentrate upon the
dynamical state of the HRS environment outside of the rich clusters.
To remedy this situation for the HRS, we have initiated a wide-field,
spectroscopic study of the inter-cluster regions, and the
initial results of our study are presented here.
Specifically, we report here on the largest scale spatial-kinematic
features found in our data.  In \S\ 2, we describe the redshift
sample, from the galaxy selection to the determination of optical
redshifts. The four primary results of the study, all relating to
large-scale kinematic features in the HRS, are presented in \S\
3. In \S\ 4 we compare our results from the HRS with studies of the
Shapley Supercluster, the largest mass concentration in the
local universe. Throughout, the following cosmological parameters are
adopted: $\Omega _m = 0.3$, $\Omega _\Lambda = 0.7$, and $H_o = 100h
= 70 $ \kms\ Mpc$^{-1}$, which implies a scale of 4.6 Mpc
degree$^{-1}$ (77 kpc arcmin$^{-1}$) at the $\sim$20,000 \kms\ mean
redshift of the HRS.

\section{DATA}
\subsection{Selection} 
The UK Schmidt Telescope (UKST) six-degree field (6dF), multi-fiber
system is uniquely suited to survey large supercluster regions in the
nearby universe. 6dF deploys 150 fibers over a circular field of
diameter 6\degr\ with a minimum required spacing between fibers of
6\farcm 7, set by the magnetic prism buttons. Light is fed from the
fibers into a fast f/0.9 CCD spectrograph \citep{par98}. Two
interchangeable field plate units allow for the simultaneous observation
of the current field and configuration of the next.  A practical limiting
magnitude for the system is b$_J$ = 17.5.  All of these attributes
taken together imply that the 6dF is most effectively used to probe
the large-scale inter-cluster environments of local superclusters,
while avoiding the more densely crowded cluster members.
Consequently, in studying the HRS our goal was to produce a catalog of
galaxies in the inter-cluster region.

Galaxy selection took place in the following manner. A 12\degr\  $\times$
12\degr\ area of the sky centered upon $\alpha = 3^h19^m, \delta =
-$50\arcdeg 02\arcmin\ was chosen for the region of observation based 
upon previously published literature \citep{zuc93}. A complete catalog
of all galaxies down to a b$_J$ magnitude of 17.5 was extracted in
four equivalent (6\arcdeg  $\times$ 6\arcdeg ) regions from the UKST
survey plates previously scanned by the 
SuperCOSMOS machine \citep{ham01}. There was also the addition of a
fifth rectangular region (3\degr $\times$ 6\degr) in the far Southern
portion to incorporate the field surrounding ACO clusters 3106 and 
3164. The galaxy classification flag assigned by SuperCOSMOS was used
for the initial sample selection. The b$_J= 17.5$ magnitude limit was
adopted as a practical limiting magnitude for the 1.2-m aperture
UKST. To avoid expending fibers on galaxies within clusters, our
original intention was to excise from the catalog all galaxies within
a 1$^{\circ}$ radius circle of sixteen ACO clusters  listed by
\cite{zuc93} as members of the HRS and intersecting our observing
region. The 1$^{\circ}$ radius exclusion corresponds to $\sim$2 Abell
radii (where 1 R$_A=$ 2 Mpc) at the mean
redshift of the HRS.  This would ensure that new spectroscopic
information relates only to the inter-cluster regions of the HRS.
However, a coding error was discovered in the program that excises galaxies
from the cluster regions only after the observations were made. 
The $\cos$($\delta$) conversion factor in the Right
Ascension (RA) coordinate, when expressed in degrees, was not included in
the calculation of angular distances of galaxies from cluster
centers. As a result, the actual excision regions are flattened in the
RA coordinate and correspondingly more flattened at higher
Declination.  The typical flattening is a factor of 1.6.
Nevertheless, the result remains that we have generated a sample that
is almost entirely comprised of inter-cluster galaxies.  

\begin{figure}
  \plotone{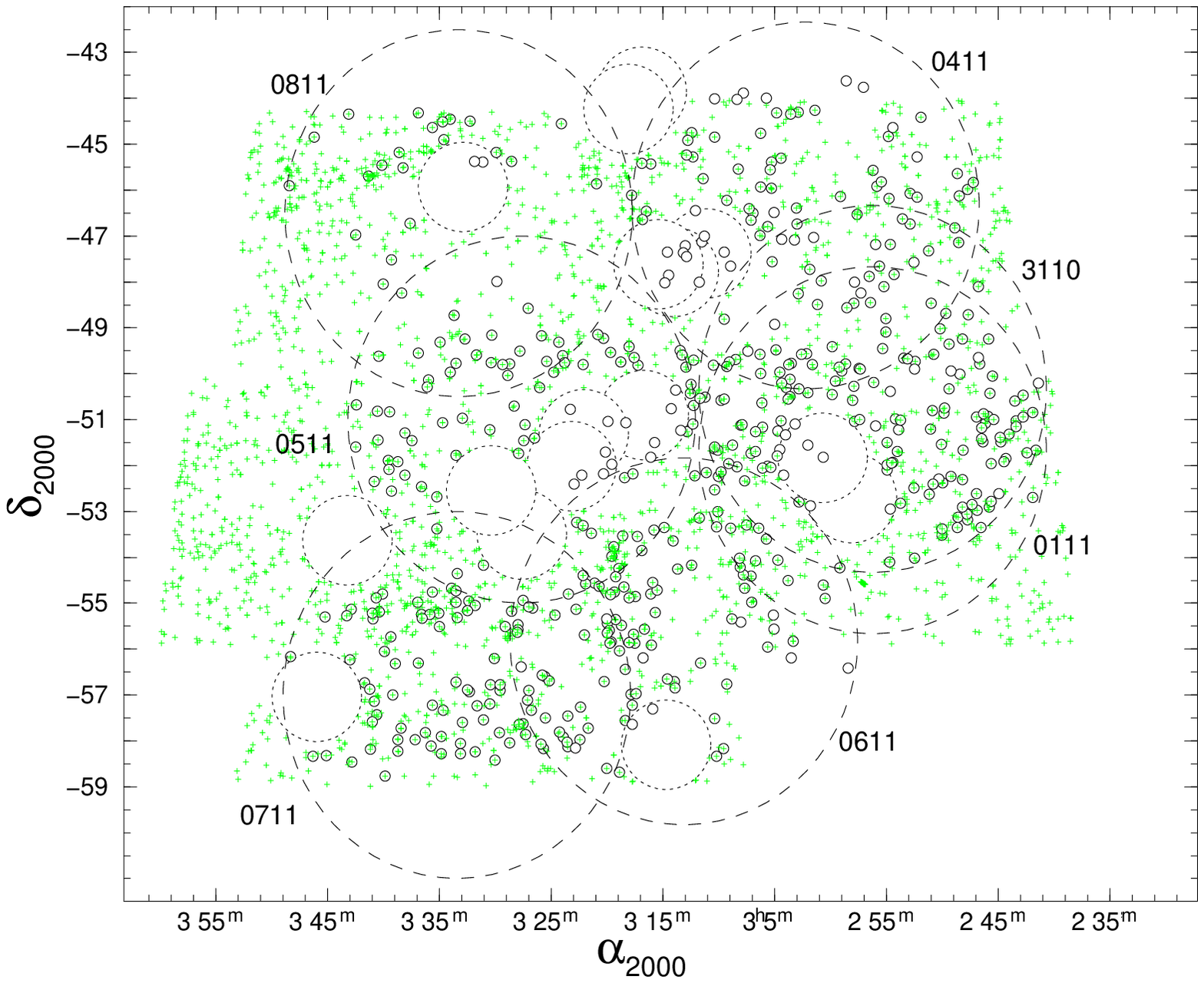}
  \caption{Observed fields in the present study as conducted by the
    6dFGS team. Crosses represent all 2848 galaxies from the
    SuperCOSMOS catalog. Note that, as described in the text,
    one degree radius regions ($\sim$2 R$_{A}$) around 16 ACO
    clusters listed as members of the HRS by \citet{zuc93} are
    excluded from the catalog.  The excised regions are shown as
    dotted circles. Small, open circles represent galaxies for which optical 
    redshifts were obtained. The 6dF r$-\theta$ positioner selects a
    6-degree diameter region from the UKST field plates, which are denoted by
    large dashed circles. Labels refer to the spectroscopic
    observations detailed in Table \ref{tb1} (column
    4). \label{f1} }  
\end{figure}

After the above constraints were applied, there remained 2848 galaxies
(Figure \ref{f1}). The maximum number of optical galaxy
redshifts that could be obtained under optimal observing conditions
was estimated at 1500. Consequently, we produced a subcatalog of 1500
targets from the original list of 2848.  This was accomplished as follows.
Galaxies in each 6\arcdeg  $\times$ 6\arcdeg\ region were assigned a
random number and then arranged in ascending order. This ordering
provides a basis for selecting an unbiased subsample from the larger
complete sample.  The numbering schemes from the individual 6\arcdeg
$\times$ 6\arcdeg\ regions were merged into a final catalog of 1500
objects with each region weighted according to the fraction of
galaxies found in that region.  That is, if 25\% of the galaxies in
the original catalog came from a particular region, the subcatalog of
1500 galaxies also contained 25\% from that region. Hence the method
preserves natural galaxy overdensities while randomly sampling the
entire extracted region. Finally, a Digitized Sky Survey (DSS) \footnotemark{}
image of each target was examined to further reduce the number
of misclassified galaxies in the sample.  

\footnotetext{The Digitized Sky Surveys were produced at the Space
Telescope Science Institute under U.S. Government grant NAG W-2166.}  

\subsection{Observations} 
\begin{figure}
  \plotone{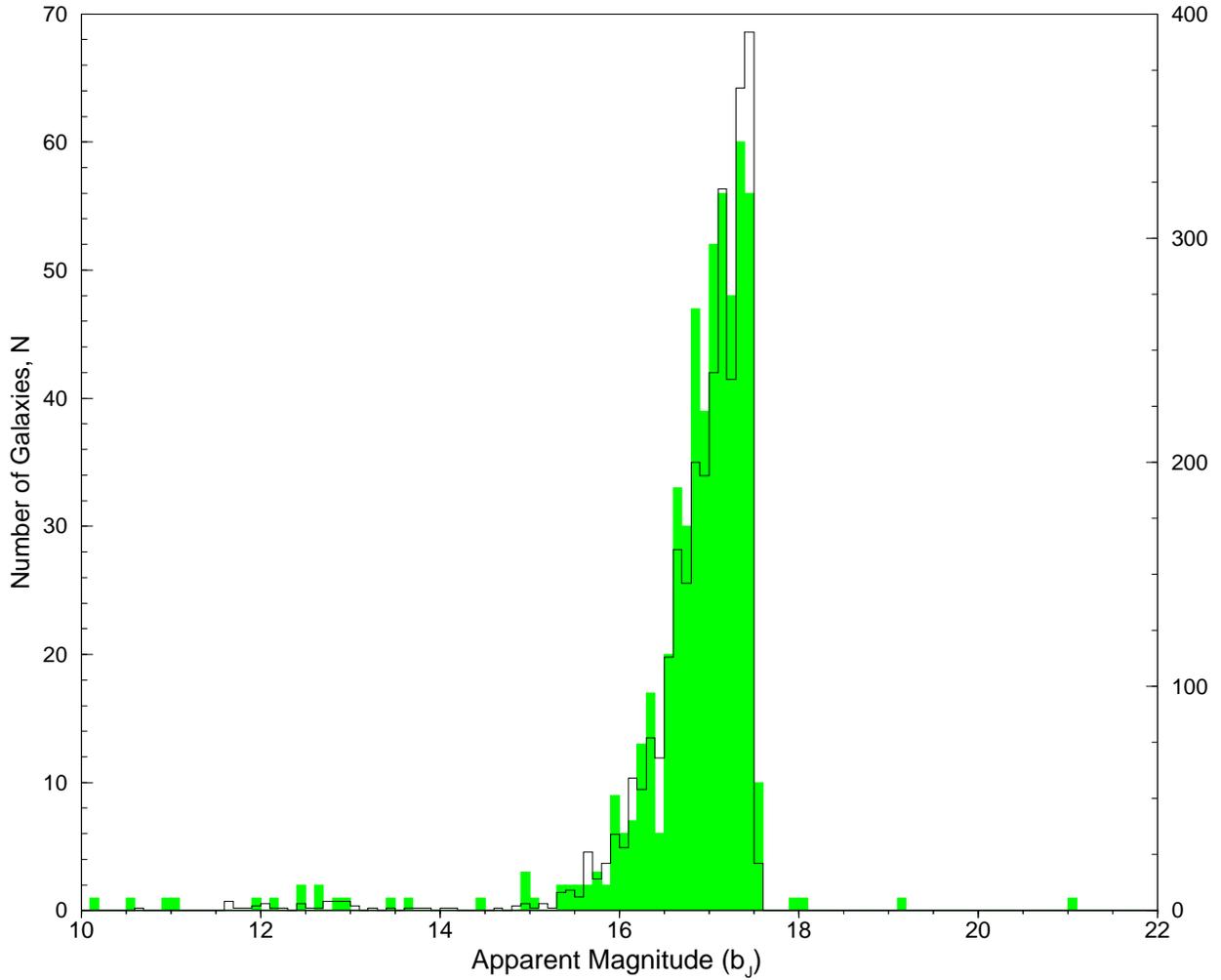}
  \caption{Histogram showing the magnitude distribution for the 6dF
  observations compared to the SuperCosmos inter-cluster galaxy list with
  limiting magnitude b$_J =$ 17.5. Filled histogram shows the
  magnitude distribution of the observed objects (547) and correlates with the
  y-axis labeled on the left-hand side. Outlined histogram shows the
  original list of galaxies (2848) after the cluster
  galaxies were removed and correlates with the labels on the right-hand
  side. \label{f2}}   
\end{figure}

Observations covering the 12\degr $\times$ 14\degr\ area in the HRS
were carried out on the 1.2m UKST of the Anglo-Australian Observatory
(AAO) in 2002 October/November. All observations were carried out as
part of the 6dF Galaxy Survey (6dFGS) program being undertaken by the AAO
\citep{wak03}. Specifically, the 6dFGS and our HRS program
observations were folded together to allow for joint execution of both
programs. When allocating fibers, the 1500 galaxies in the study were
given highest priority within the 6dFGS for the selected fields of
observation. However, whenever a 6dF fiber became unassigned due to a
conflict with the fiber selection from another target galaxy, the
fiber was then reassigned to a target from the 6dFGS lists. The blue
magnitude limit for the 6dFGS is 16.75 (i.e., b$_J 
<$ 16.75), hence there is considerable overlap between
our target lists and the 6dFGS. Over all the observed fields,
approximately 70\% of all targets were taken from our original list of
1500 galaxies. Noticeable from Figure \ref{f2}, our
observed galaxy magnitude distribution closely follows the magnitude
distribution of the post-extraction HRS area of 2848 galaxies.  Due to
the brighter limiting magnitude of 6dFGS, we have slightly less
proportional coverage at our faint limit.  In addition, a few very
faint objects were included as part of the 6dFGS, which again can be 
seen in Figure \ref{f2}.  Finally, a small number of
6dFGS objects lie within our 1\degr\ excision radii around clusters,
which is evident in Figure \ref{f1}. 
 
Observations and reductions were carried out along standard 6dFGS
procedures, which are briefly summarized in the next section. Eight
nights were allocated to this project by the 6dFGS team, 
but three were adversely affected by weather (Table \ref{tb1}). We used a
combination of the 580V and 425R volume-phase holographic transmission
gratings to optimize spectral coverage. This procedure yielded an
instrumental resolution of 4.9 \AA\ (580V) and 6.6 \AA\ (425R), while
covering the wavelength range 3900 $-$ 7600 \AA, i.e., from
[OII]$\lambda$3727 through H$\alpha$ over the HRS redshift
range. Exposure times for each grating are listed in Table
\ref{tb1}. HgCdNe arc and quartz flat exposures were carried out
before and after primary fields.  

\subsection{Reductions}
547 usable galaxy spectra were obtained from the eight nights allocated
(Table \ref{shorttb2}). In Figure \ref{f1}, individual field
centers are labeled and shown in reference to the survey area. Altogether, 100
fibers were operational during our sequence of observations. With 9
fibers donated to sky, this leaves a total of 91 possible galaxy
redshifts per imaged field. Night 7 with the 580V grating was not
reduced due to a telescope focus error, so redshifts were obtained for
only 25\% of the 0811 field (Table \ref{tb1}). Although the signal-to-noise
ratio was relatively low in many of our spectra ($<$ 10), over 95\%
yielded reliable redshifts (excluding 0811). Due to 6dFGS priorities
and galaxy overcrowding, redshifts were obtained for some galaxies not 
originally included in our source lists. There remained 3 Galactic
stars and 27 objects with unusable spectra in the sample.   

The automatic 6dF data reduction ({\tt 6DFDR}) package completes the
following steps directly after observation: debiasing, fiber
extraction, cosmic-ray removal, flat fielding, sky subtraction, and
wavelength calibration \citep{jon04}. As a final step, the post-{\tt
6DFDR} files from each exposure were co-added into single spectra.  

\subsection{Redshift Determination}
Methods for the determination of galaxy redshifts fit into three basic
categories depending on their spectral characteristics: absorption,
emission, and those spectra containing both absorption and emission
features. For spectra exhibiting absorption features, the  
IRAF \footnotemark{} based cross-correlation package, {\tt rvsao}, was
utilized to determine radial velocities against four template spectra: two 
stellar spectra obtained from the Coud\'e Feed spectral library 
\citep{jon99} and two spectra obtained from the sample (a Galactic   
star and a nearby galaxy whose redshift was also determined by {\tt
  rvsao}). 

\footnotetext{Image Reduction and Analysis Facility (IRAF) is written and
supported by the National Optical Astronomy Observatories (NOAO)
and the Association of Universities for Research in Astronomy (AURA),
Inc. under cooperative agreement with the National Science Foundation.} 

The method of determining redshifts for emission-dominated galaxy
spectra was completed in two steps. First, JAR and MCF independently
measured wavelength centers for each detectable spectral line via Gaussian
fitting then determined its redshift. Second, each emission line was
assigned a weight by MCF based upon the sharpness of the line
and the surrounding noise level. The assigned weight was based upon
a 5 point scale, where a ``5'' denoted a peak height greater than
three times the FWHM with minimal background. For expected emission
lines that were faintly detectable from the background, a
weighting of ``1'' was assigned. This appropriately distinguished
between emission lines with robust redshift determinations from those 
compromised by noise. Redshifts were averaged for galaxy spectra
exhibiting both strong emission and absorption features. Whenever there was a
discrepancy of $\Delta cz > 100$ \kms\ between the two methods,
preference was given to the emission line value. As a last step,
heliocentric corrections were applied to all redshifts.

\subsection{Coverage}
\begin{figure}
  \plotone{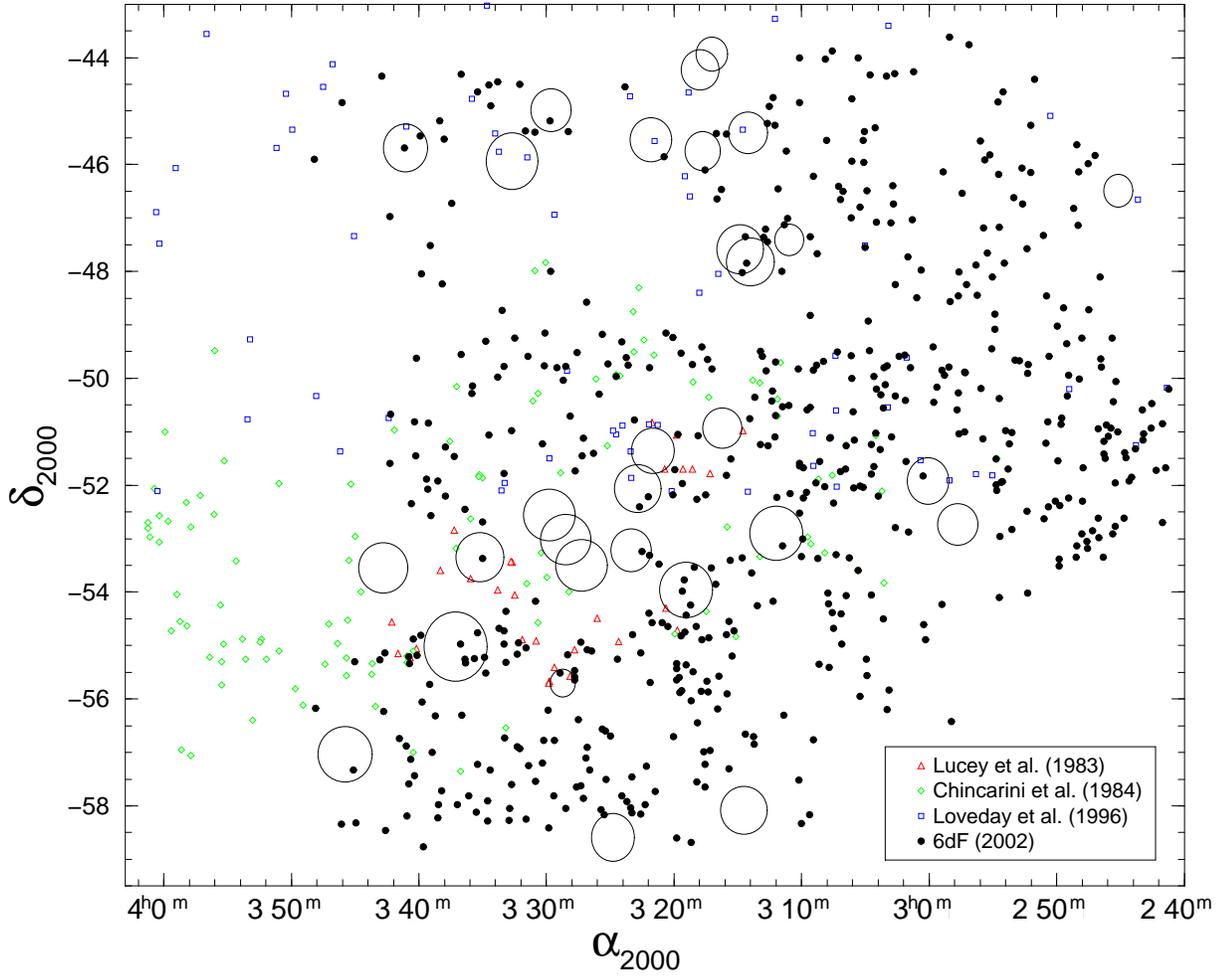}
  \caption{The HRS region under study displaying both 6dF and previous
  inter-cluster redshifts. Previously observed galaxies are plotted with
  different symbols to show the increased amount of information with our 
  6dF study. Redshifts from our 2002 study are shown as filled
  circles. Clusters in the
  observing region with known redshift (Table \ref{tb3}) are shown
  as large open circles with radii of 2 Mpc (1
  R$_{A}$). \label{f3}} 
\end{figure}

Outside the previously determined cluster areas that were excised,
there were 2848 potential targets selected by SuperCOSMOS (galaxies
with b$_J <$ 17.5). It was determined from a comparable sub-sample
selection that $\sim$15\% of the targets labeled as `galaxies' by
SuperCOSMOS were actually stars. Therefore, the completeness of the
survey is 547/2420, or 23\%. The optical redshifts obtained in this
survey more than double the previously published information regarding
the proposed supercluster field (Figure \ref{f3}). Previous
inter-cluster observations were limited both spatially (primarily
focused in the Southeast portion of the supercluster) and also by
magnitude. Overlap with previously observed galaxies was not intended,
but in general, most of the 6dF redshifts are consistent with previous
estimates within the combined errors.

It is noticeable from Figure \ref{f1} that the coverage is
not uniform over the original 12\degr $\times$ 14\degr\ 
area. In fact, the total area covered by the observations is more
accurately 9\degr $\times$ 14\degr. Furthermore, the galaxies in the Western
portion are more heavily sampled than those in the East. This
non-uniformity is primarily a result of the weather problems coupled
with the competing demands of both HRS and 6dFGS surveys when selecting
field centers for the observations. Although the mean completeness is
23\%, the field centers in the Western portion are sampled closer to
28\% completeness, while the Eastern field centers are at
$\sim$22\%.

\section{RESULTS}
\subsection{Kinematic Extent of the HRS}

We begin by briefly considering the kinematic extent of the HRS.  While a
supercluster clearly will not be in a state of dynamical equilibrium with
well-defined boundaries, we establish provisional kinematic limits
from previous studies of the Horologium-Reticulum (HR) cluster
population. Specifically, two separate studies applied a
friends-of-friends analysis to the ACO clusters within the HR region
and identified between 18 and 24 clusters as related to the HRS
\citep{zuc93,ein01}. Within our region of observation, seventeen ACO
clusters were combined from these two catalogs (column 4, Table
\ref{tb3}). The mean redshift of these clusters is 19,900 \kms,
with a dispersion $\sigma$ of 2300 \kms. We define the ``kinematic
core'' of the HRS to be roughly bounded by the FWHM of the observed
redshift distribution of ACO clusters, namely 5400 \kms. When rounded
to the nearest 500 \kms, we determine the core of the HRS to be
between $cz$ of 17,000 and 22,500 \kms.   

The above kinematic extent of $\sim$5500 \kms\ is basically consistent 
with the fact that the \citet{zuc93} and \citet{ein01} analyses find
the HRS to contain $\sim$20 major galaxy clusters. Assuming a mean
cluster mass of 10$^{15}\,M_{\sun}$ and the cosmological parameters
stated in \S1, we calculate the spherical Hubble Flow volume required
to contain a mass of 20 such clusters. The calculated diameter of that
volume (95 Mpc) does indeed correspond to a velocity spread of
$\sim$6500 \kms, i.e., similar to our defined kinematic limits. 
 
\begin{figure}
  \plotone{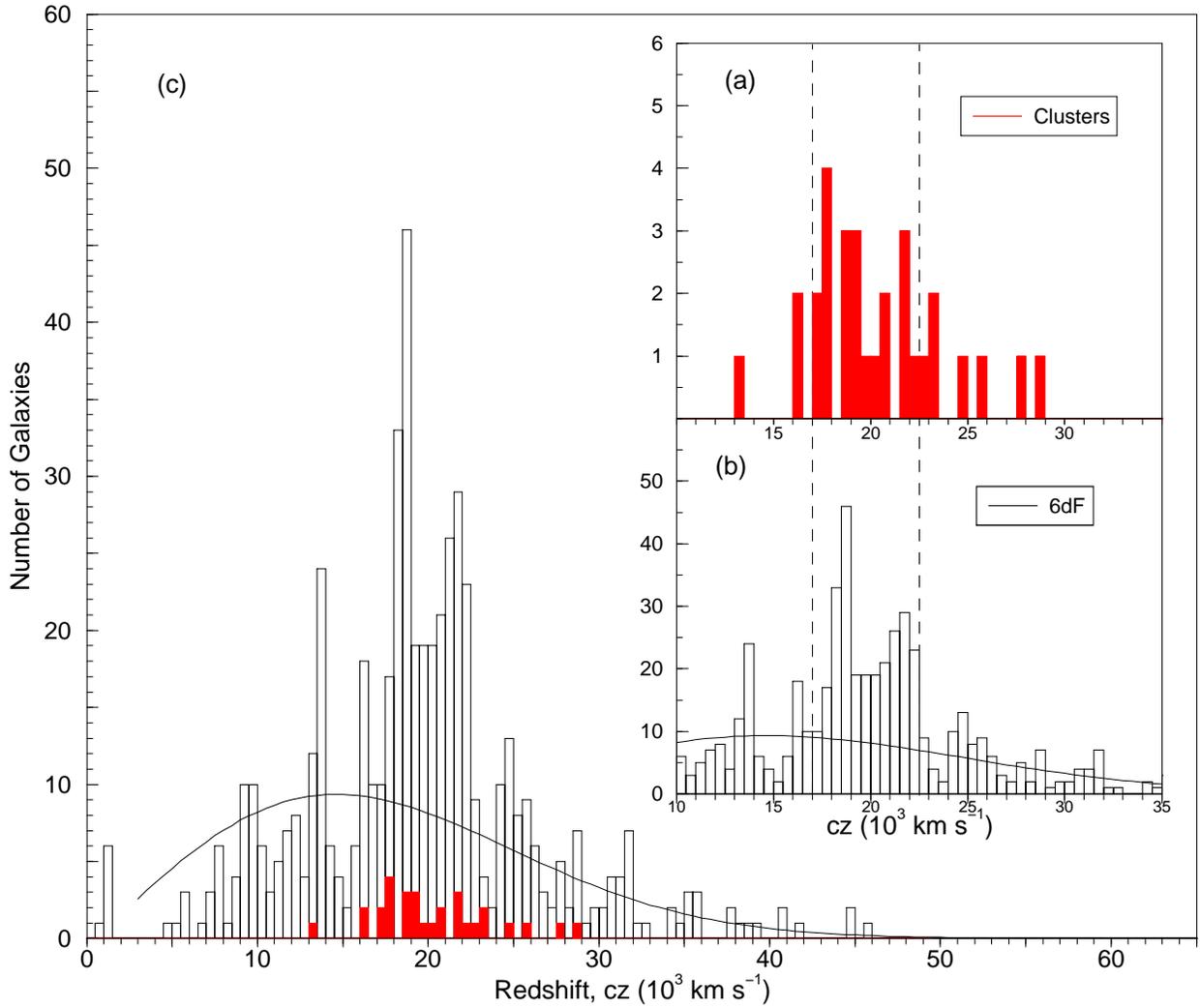}
  \caption{Redshift histograms of the 6dF inter-cluster galaxies (open) and
   the clusters with known redshifts (filled). Panel (a): Cluster redshifts
   from Table \ref{tb3}. Panel (b): Redshifts for inter-cluster galaxies 
   covering the same range as clusters in panel (a). Dashed lines in both
   inset histograms represent the kinematic ``core'' discussed in the text. In
   panel (c), we show the entire redshift histogram for the
   inter-cluster galaxies with the clusters overlaid. Solid line shown
   in both inter-cluster galaxy histograms is the expected number of
   counts for a smooth, homogeneous distribution. The redshift bin
   size is 500 \kms. \label{f4}}  
\end{figure}

\begin{figure}
  \plottwo{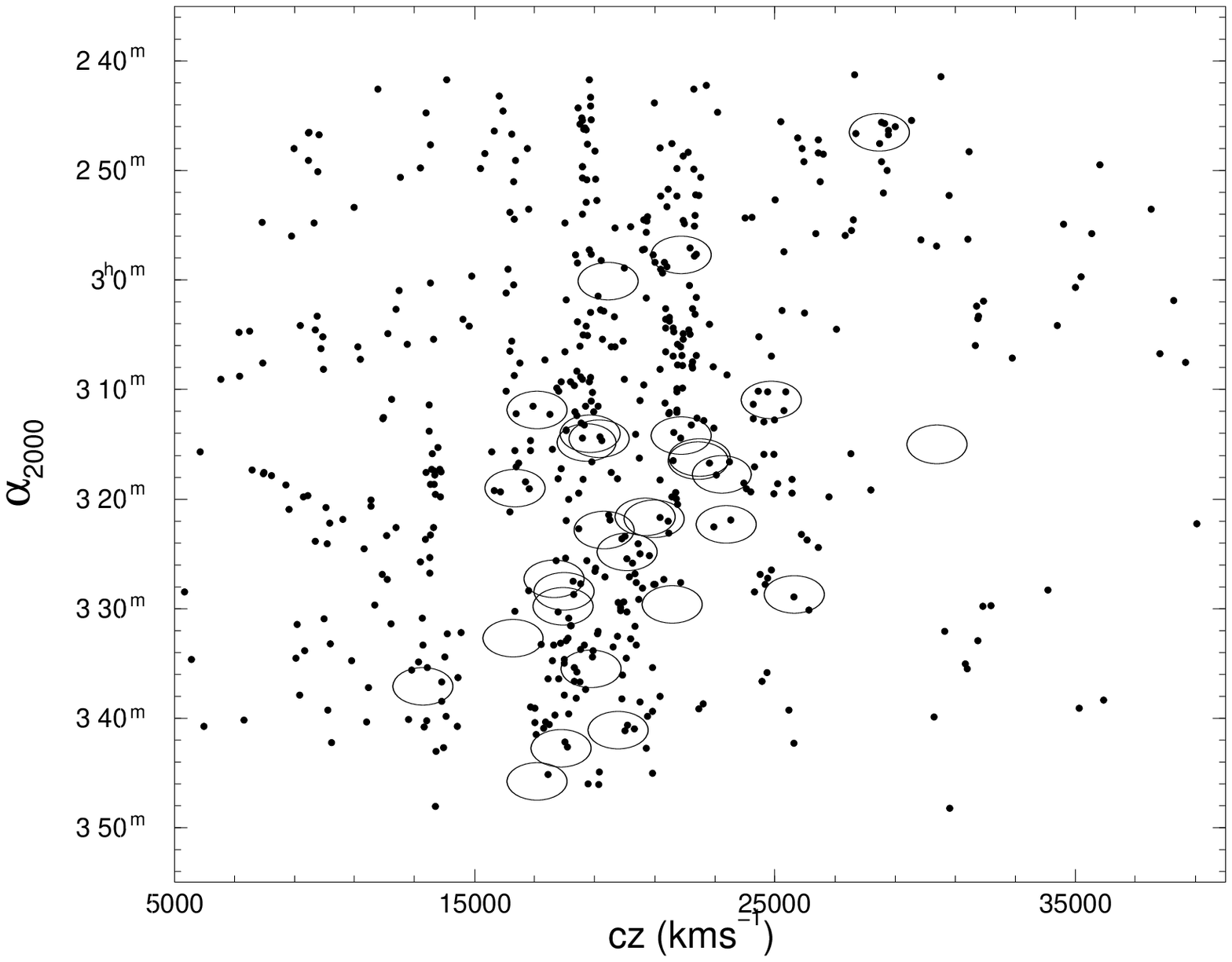}{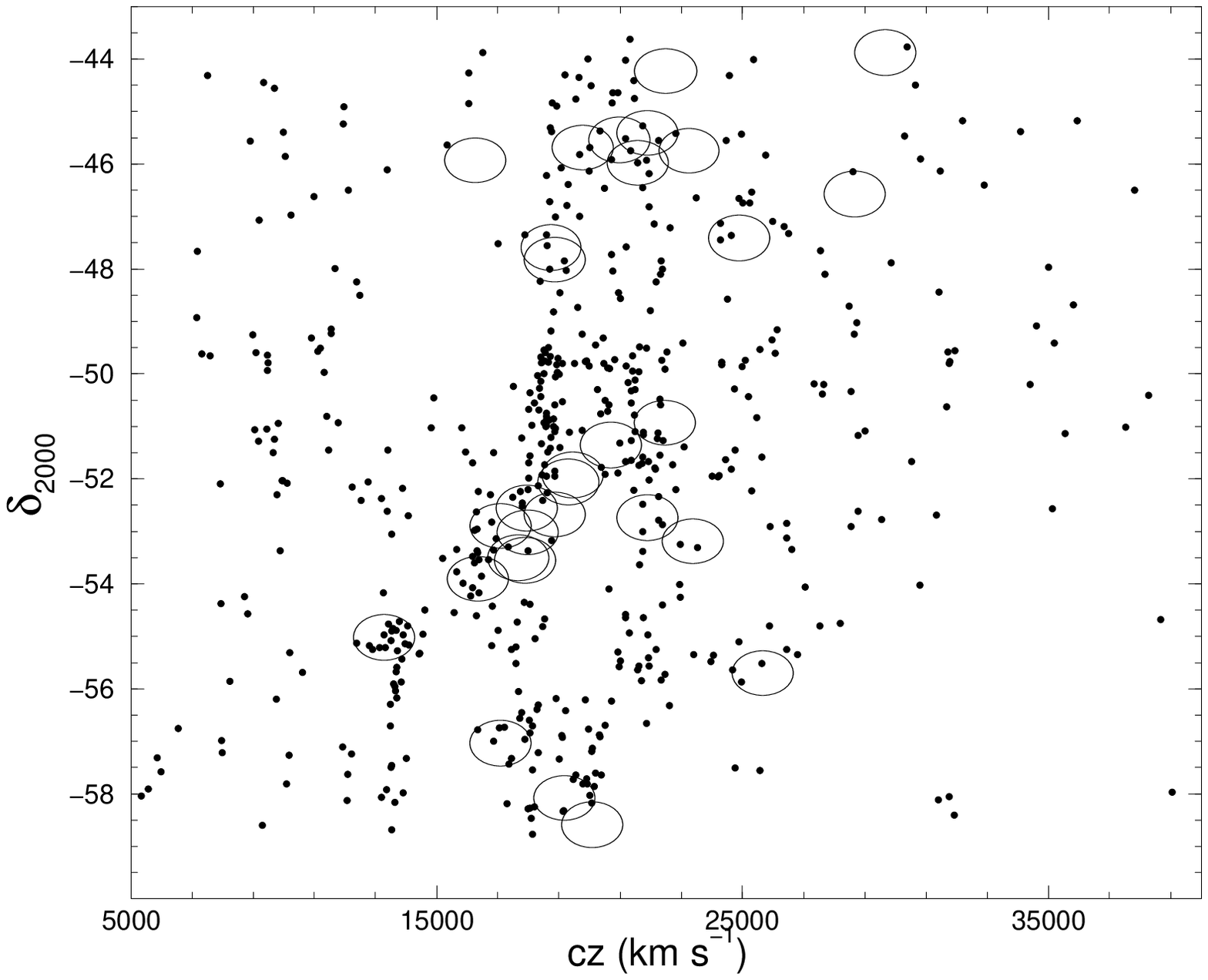}
  \caption{Coordinate-redshift plots for the 6dF galaxies. Left panel:
  $\alpha -cz$. Right panel: $\delta -cz$. Clusters in
  Figure 2 are shown as ellipses with an estimated velocity dispersion of
  1000 \kms\ (horizontal axis) and a vertical axis of 4
  Mpc (2 R$_A$) at the mean HRS redshift (20,000 \kms). \label{f5}}   
\end{figure}

The adopted boundaries for the HRS are examined with respect to
both the distribution of cluster redshifts and the distribution of
6dF inter-cluster galaxy redshifts in Figure \ref{f4}. For
inset (a), all galaxy clusters with known redshifts in the region
(Table \ref{tb3}) are plotted, including the seventeen ACO clusters
considered above. Although the redshift histogram for the
inter-cluster galaxies is clearly clumped into several redshift
concentrations, the main concentration of galaxies ($\sim$48\% of the
sample) lies within the selected HRS kinematic 
boundaries.  As shown in Figure \ref{f4}(b), the HRS
kinematic core is bordered by a depletion in galaxy numbers both at lower
(14,000 $-$ 16,000 \kms) and higher (22,500 $-$ 24,000 \kms) redshift
ranges. While the higher redshift limit to the HRS near cz = 22,500
\kms\ is quite well defined, the lower redshift limit is less clearly
defined. Specifically, there is a clump of galaxies present between 
16,000 and 17,000 \kms, which is not included in our definition
of the HRS ``core''. The nature of the inter-cluster galaxies in this
redshift regime is further clarified in Figure \ref{f5},
where we plot coordinate versus redshift in both $\alpha$ and
$\delta$.  Note that the galaxies between 16,000 and 17,000 \kms\ in
redshift are highly concentrated in $\delta$ at
$\sim-$54\arcdeg. These same galaxies are more substantially spread in
$\alpha$, although confined to the Western side of the HRS. We return
to this component of the HRS in subsequent sections.  

\subsection{Inter-cluster Galaxy Overdensity}

Our extensive new redshift database allows
us to calculate the mean galaxy overdensity in the {\it inter-cluster}
regions of the HRS.  The expected galaxy counts for a uniform distribution 
are based on estimates of the local galaxy luminosity function (LF).
To facilitate the comparison between the HRS and the SSC, we follow as
closely as possible the methods described by \citet{dri04}.
Specifically, the expected number of galaxy counts as a function of
redshift and limiting magnitude are calculated from the same
\citet{met91} galaxy LF as used by \citet{dri04} in their calculation
of the overdensity in the SSC. The resulting function is plotted as a
solid curve in Figure \ref{f4}, with an assumed limiting
magnitude of b$_J$ = 17.5. To calculate the expected number of
galaxies within the HRS redshift and angular limits, we adopted the
previously established redshift limits of 17000 $-$ 22500 \kms, and
assumed the 9\arcdeg\ $\times$ 14\arcdeg\ = 126 deg$^2$ areal coverage of our
survey. However, the latter figure required reduction to 107 deg$^2$,
due to the areas excised around clusters.  Finally, we observed only
23\% of the total number of galaxies brighter than b$_J$ of 17.5
within the 107 deg$^2$.  Of the observed galaxies, $\sim$48\% fall
within the redshift limits of the HRS. Taking these factors into
account, we arrive at a mean density of 2.4:1 for the inter-cluster
regions of the HRS (assuming that light traces mass). In following a
common definition of the galaxy overdensity, we find that
$\bar{\delta} = $ 1.4, where $ \bar{\delta} = (\rho_{HRS} -
\bar{\rho}) / {\bar{\rho}}$. Given the 
rather uncertain redshift and angular boundaries of the HRS, as well
as uncertainty in the shape and normalization of the local LF, we
estimate an uncertainty of $\sim$25\% in the overdensity.

\subsection{Large-Scale Redshift Trend}

\begin{figure}
  \plotone{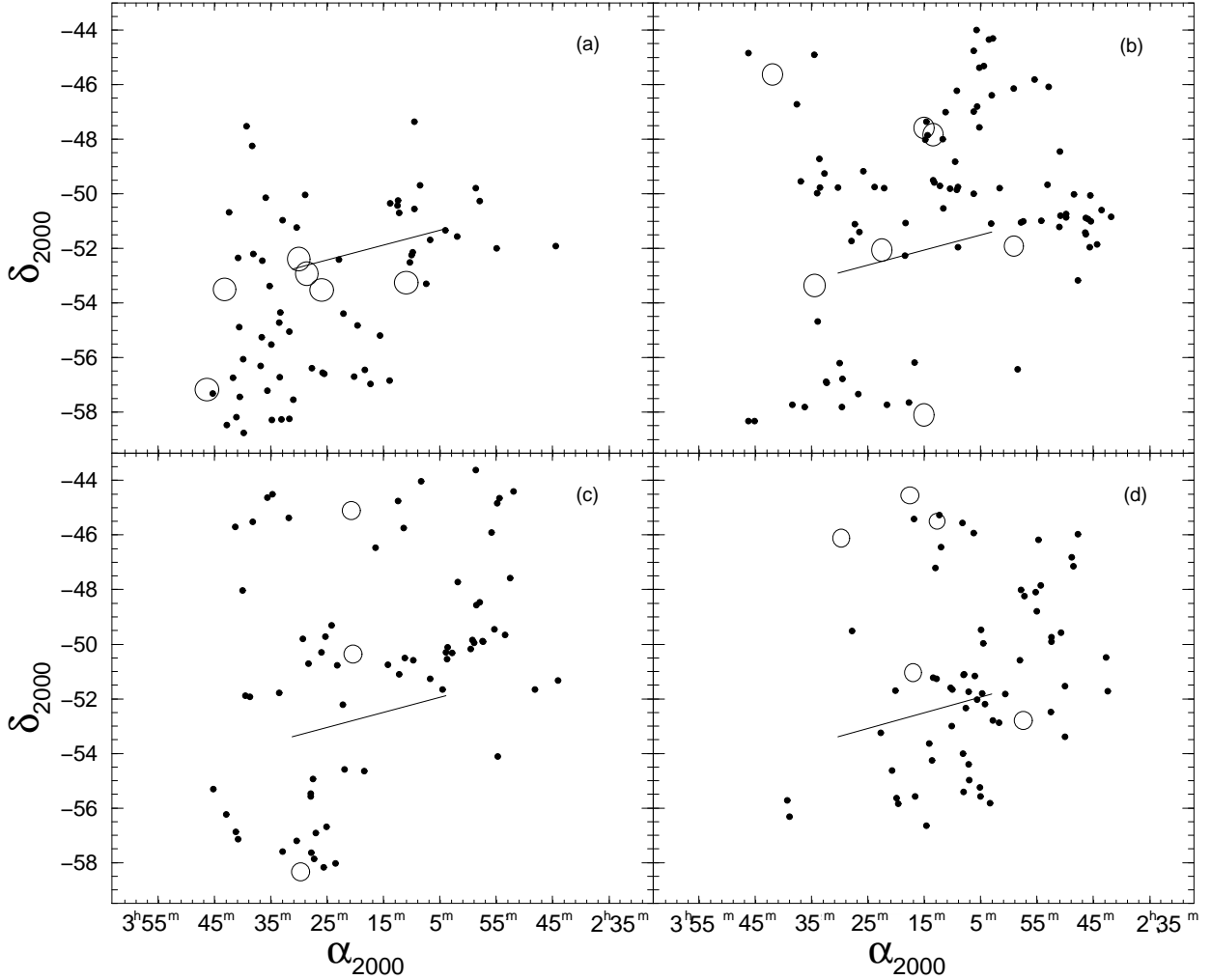}
  \caption{Redshift slices are plotted for the 6dF data in the range
  of the HRS.  Each panel covers a 1500 \kms\ redshift slice.
  Individual galaxies are plotted as small filled circles.
  Clusters from Figure 2 are also included in their respective
  redshift ranges. The short solid lines in each panel show the
  best-fit axis from the spatial-cz correlation analysis (PA$=-$80\degr).
  Only a short line is drawn because of the curvature produced by
  the non-equal area of the conventional $\alpha - \delta$ coordinate
  projection. Panel (a) 17,000 $-$
  18,500 \kms, Panel (b) 18,500 $-$ 20,000 \kms, Panel (c) 20,000 $-$
  21,500 \kms, Panel (d) 21,500 $-$ 23,000 \kms.\label{f6}}  
\end{figure}

Having examined the overall redshift histograms for both clusters and
inter-cluster galaxies, we now utilize two-dimensional redshift slices
of our 6dF data as a further means of assessing the dynamical state of the
HRS.  In Figure \ref{f6}, we present a sequence of redshift
cuts through the kinematic extent of the HRS, each cut containing a
redshift bin size of 1500 \kms. An examination of Figure
\ref{f6} gives the 
impression of a systematic trend between spatial position in the HRS
and redshift. Specifically, we note that galaxies in panel (a) (17,000
$-$ 18,500 \kms) appear preferentially located in the South and East,
while the galaxies in panel (d) (21,500 $-$ 23,000 \kms)
preferentially populate the West and North. In other words, there
appears to be a trend of systematically increasing redshift along a
principal axis in the HRS that extends from the Southeast to the
Northwest end of the supercluster. 

To quantify the significance of a large-scale redshift trend with 
spatial position in the HRS, we conducted a correlation
analysis as a function of position angle (PA) on the sky.  To begin,
we selected the center of the HRS to be at ($\alpha = 3^h16^m, \delta  =
-52\arcdeg$) and assumed the principal axis of the HRS to be aligned
along the West-East direction.  Each galaxy was projected onto this
principal axis, and we defined the {\it S}-coordinate to be the
projected angular position of the galaxy along the assumed principal
axis. Furthermore, the {\it S}-coordinate was defined to run  
negative to positive from West to East. A linear regression analysis
was carried out between the redshift and the {\it S}-coordinate, which
yields both the correlation coefficient, {\it R}, and the likelihood
that the null hypothesis ({\it no} correlation between redshift and
projected {\it S} position) is correct (Bevington 1969). We repeated
the correlation analysis at 5\arcdeg\ increments in position angle
(PA) of the assumed principal axis over the full 180\arcdeg\
range. When the assumed principal axis is running from SE
to NW, positive {\it S} values are in the NW. In the same way, when
the assumed principal axis is running from SW to NE, positive
{\it S} values in the NE. The correlation analysis was completed both
for the 263 6dF inter-cluster galaxies with redshifts between 17,000
and 22,500 \kms\ and for the 21 clusters with mean redshifts over the
same interval.  

For both clusters and inter-cluster galaxies, we find that the null
hypothesis is rejected at probability (P) levels of P$<1\%$. At
certain PAs, the plots of $R$ versus PA show a broad peak over an
interval of 20 \hash\ 40\arcdeg. 6dF galaxies show the highest
correlation coefficient ({\it R} = 0.3) and lowest probability for the
null hypothesis (P$<$10$^{-6}$) at a PA $\approx$ $-$80\arcdeg\ (as
measured East from North).  The clusters also show a significant
correlation, with a peak at a PA of $-$50\arcdeg.  In Figure
\ref{f7}, we show the projected {\it S} position plotted
versus redshift for all inter-cluster galaxies and clusters with
redshift between 17,000 and 22,500 \kms\ at a PA of $-$80\arcdeg. The
expected cluster peculiar velocities are suppressed for clarity of the
inter-cluster galaxy distribution. The linear regression fit for the
inter-cluster galaxies is plotted as a solid line.  Note that the best
fit line actually passes through a zone of low galaxy density; this is
examined further in \S3.4.  

We are now in a position to revisit the substantial population of
galaxies from 16,000 to 17,000 \kms. To determine whether or not these
galaxies are associated with the large-scale redshift trend, we
repeat the correlation analysis as a function of PA now expanding
the redshift range to 16,000 $-$ 22,500 \kms. When the galaxies between
16,000 and 17,000 \kms\ are included, the correlation coefficient is
weakened ({\it R} = 0.2), and the probability of no correlation
increases to P$\sim$10$^{-3}$. The deviation of the 16,000 $-$  
17,000 \kms\ galaxies from the redshift trend is evident in
their spatial segregation (especially in $\delta$). We display the
spatial location of these galaxies on the sky in Figure
\ref{f8}, where the galaxies from 16,000 \hash 18,000 \kms\
are separated into 1000 \kms\ slices. Although the galaxies from
16,000 \hash 17,000 \kms\ only represent $\sim$5\% of the total
population, the spatial segregation of this clump does not follow the
overall trend of the higher redshift galaxies in the HRS and provides
a significant lever arm by which the best fit correlation axis is
altered. In short, while the spatially-localized galaxies from 16,000
\hash\ 17,000 \kms\ may reside within the HRS, they do not appear to
follow the large-scale redshift trend established by the clusters and
inter-cluster galaxies over the range 17,000 \hash\ 22,500 \kms.  

\subsection{Bi-Modal Kinematics of the HRS}

\begin{figure}
  \plotone{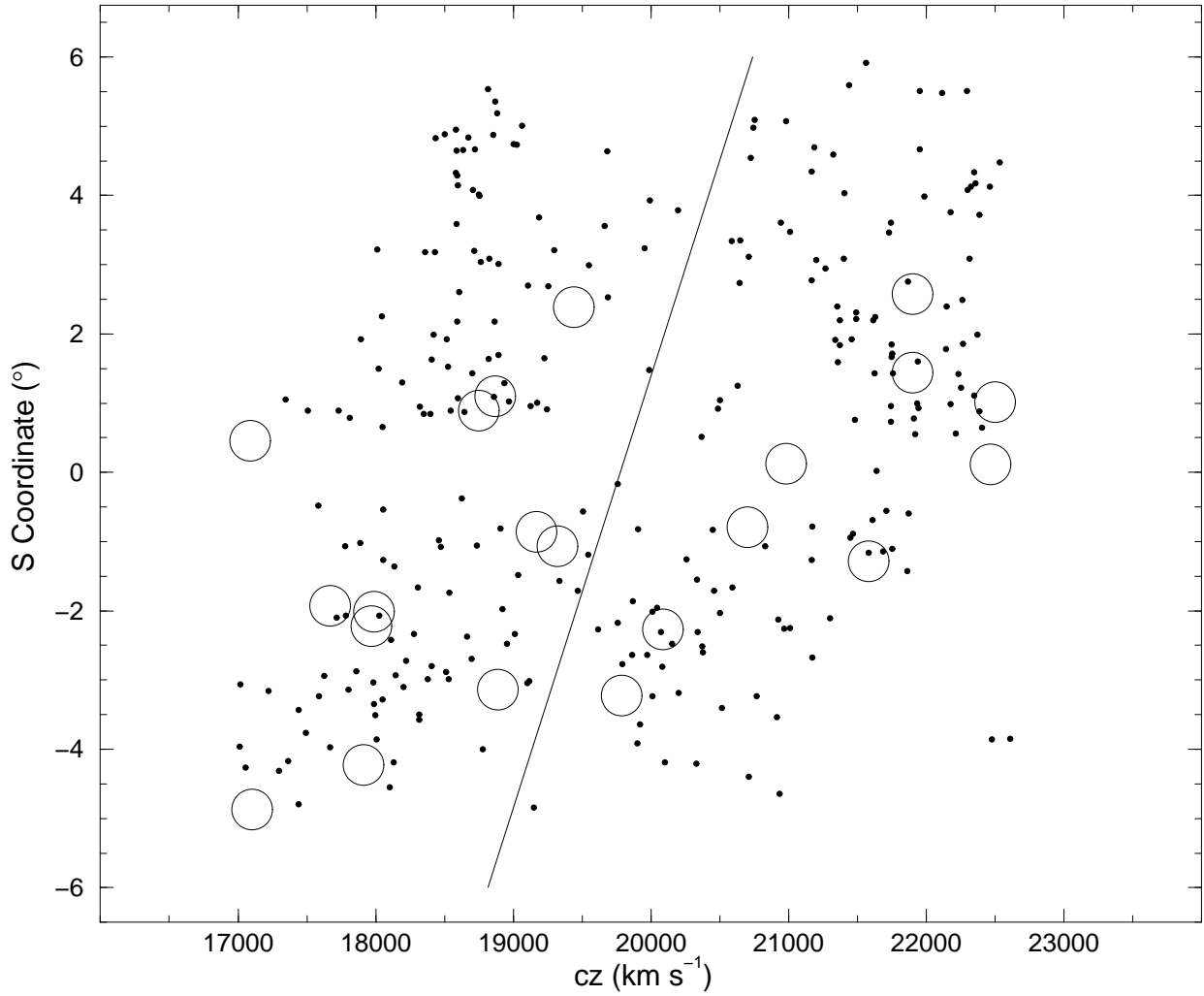}
  \caption{ Projected angular {\it S}-coordinate (see text) is plotted
  versus redshift for 6dF galaxies between 17,000 and 22,500 \kms. 
  The position angle (PA) of the principal axis of projection is at
  $-$80\arcdeg\ (as measured East from North), with positive {\it S}
  values in the NW. Individual galaxies are
  plotted as small filled circles, while open {\it circles} represent
  clusters in the region. The best fit linear regression is plotted as
  a solid line. \label{f7}}   
\end{figure}

\begin{figure}
  \plotone{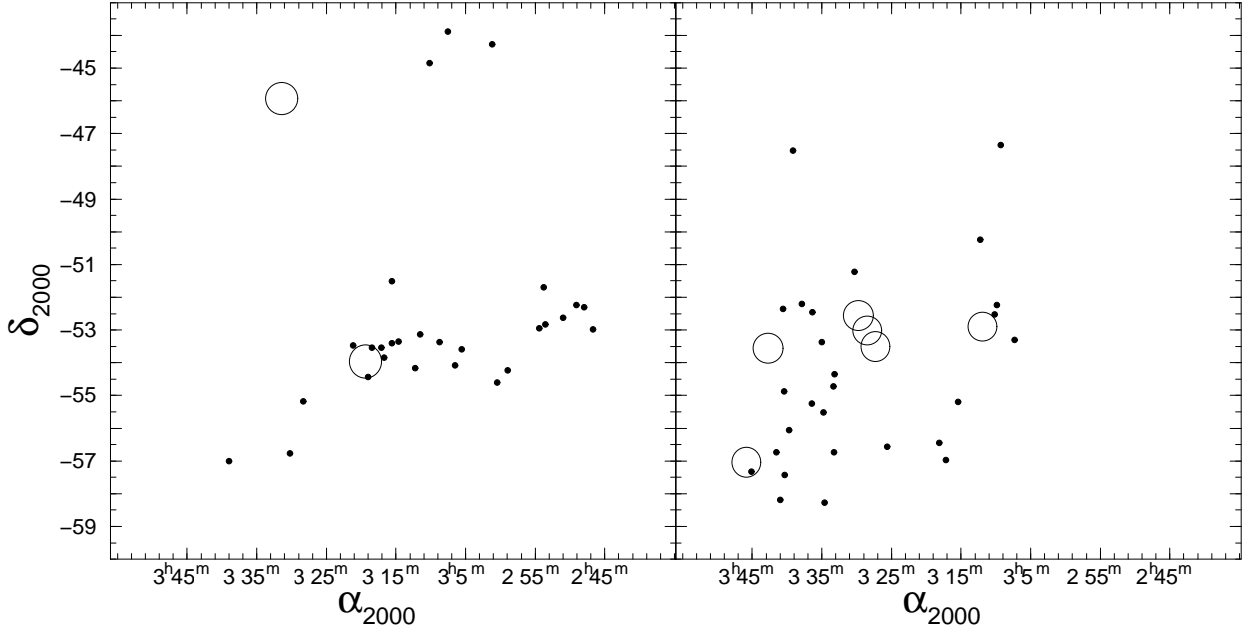}
  \caption{Separation of the 16000-18000 \kms\ redshift slice
  into low- and high-redshift bins. The 16,000 $-$ 17,000 \kms\
  and 17,000 $-$ 18,000 \kms\ galaxy populations are plotted in the
  left and right panels, respectively. Symbols as in Figure
  \ref{f3}. \label{f8}}    
\end{figure}

\begin{figure}
  \plottwo{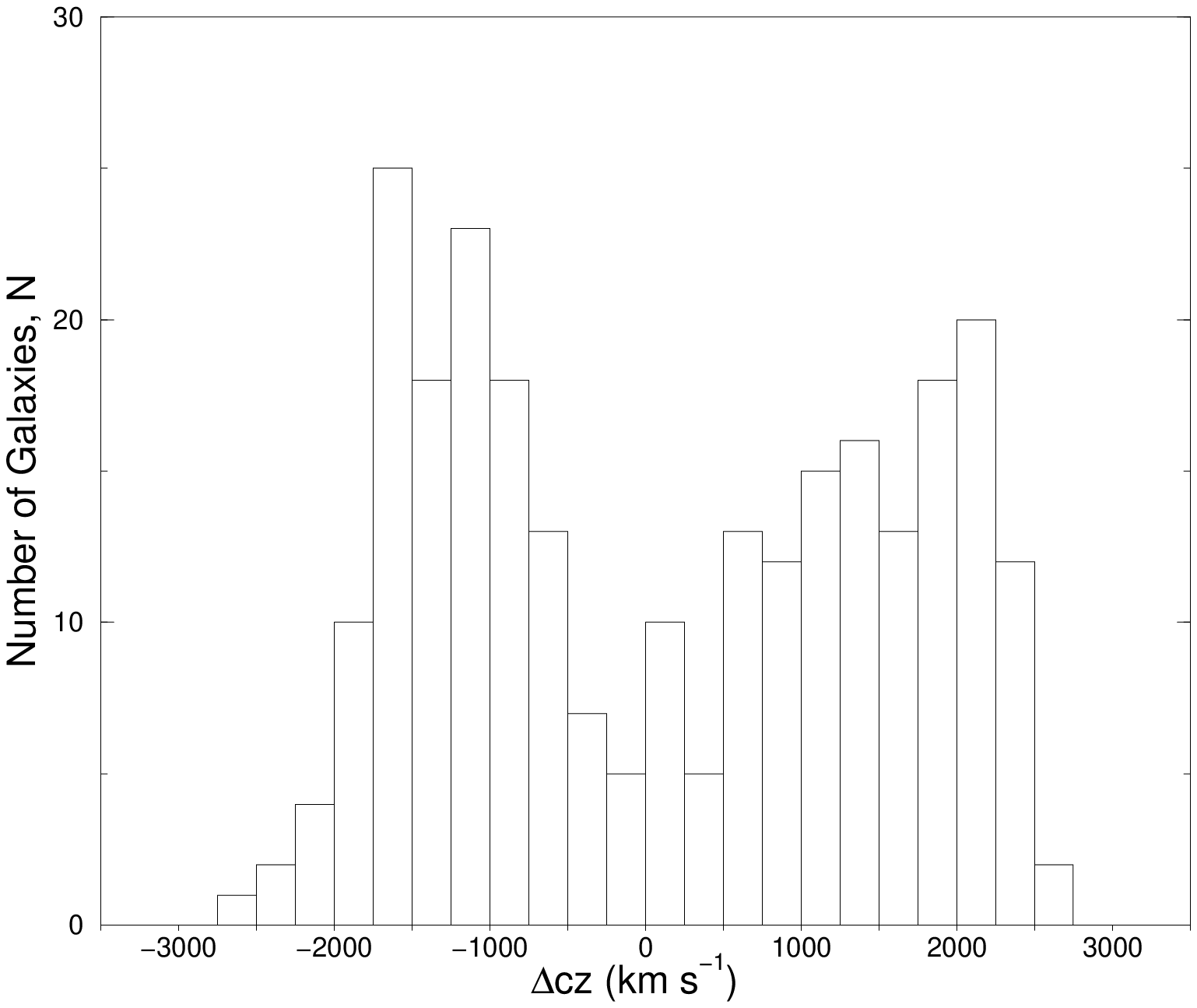}{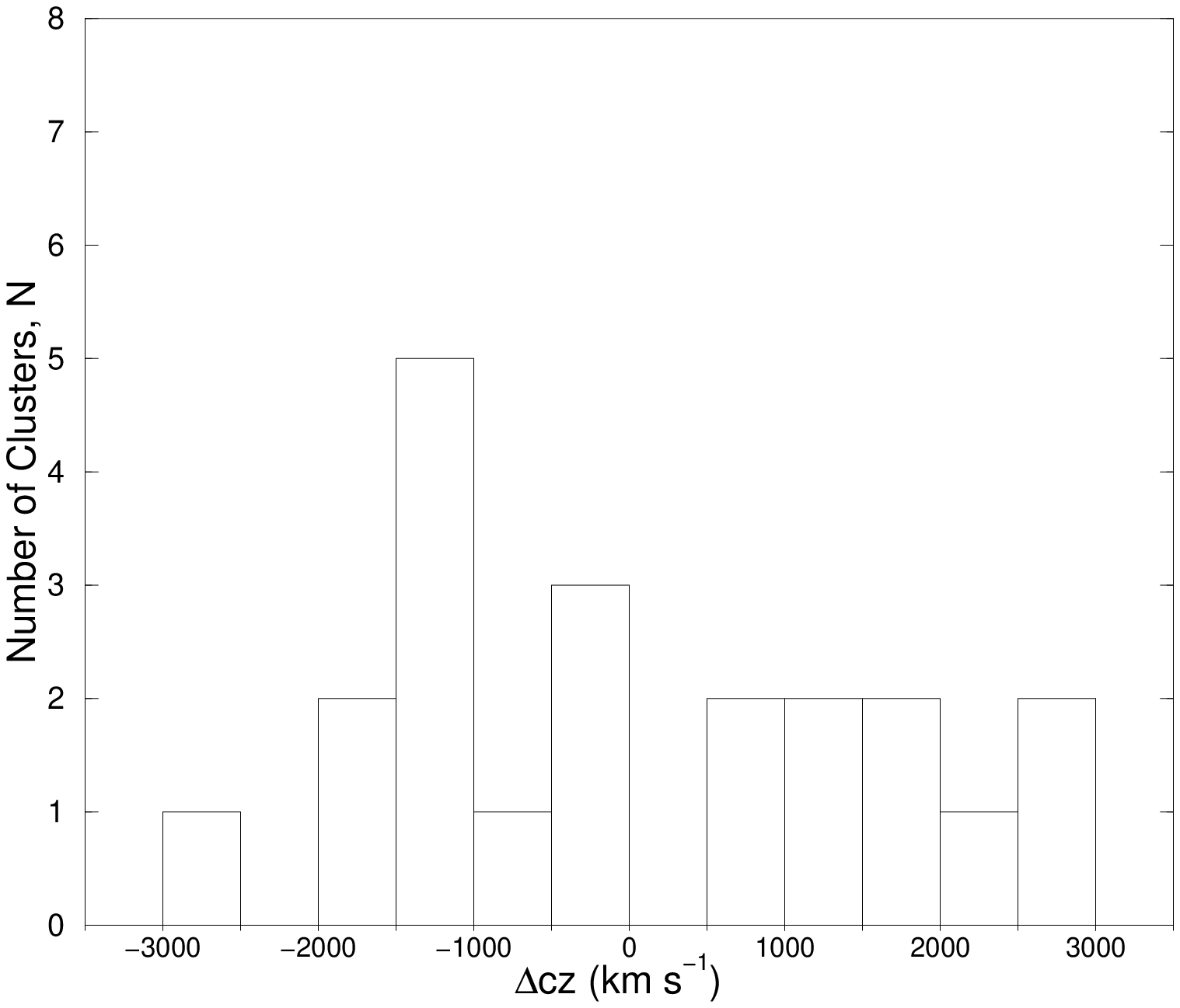}
  \caption{ Residual redshift histograms for galaxies and clusters within
  17,000 $-$ 22,500 \kms.  The residual redshift is measured relative
  to the linear regression line that represents the best fit between
  projected {\it S} coordinate and redshift for an assumed principal axis
  oriented at PA$=-$80\arcdeg, i.e., the best fit line in Figure
  \ref{f8}. Left panel: 6dF galaxies, 250 \kms\ bin size ;
  Right panel: Clusters of known redshift, 500 \kms\ bin
  size. \label{f9}}     
\end{figure}

As noted above, the best linear fit between projected {\it S} coordinate and
redshift actually runs through a zone of low galaxy density (Figure
\ref{f7}). The implication is that the HRS has a bi-modal redshift
distribution, i.e., the HRS kinematic extent consists of two major
components in redshift.  The redshift bi-modality of the HRS is most clearly
observed by fitting and removing the systematic spatial-redshift trend at
PA $=-$80\arcdeg, then plotting the histogram of residual redshifts,
as shown in Figure \ref{f9}. Fitting each component of the
histogram with a Gaussian reveals that the overall number of galaxies
is roughly equal in the two components, which are separated by
$\sim$2500 \kms\ (35 Mpc). However, the FWHM of the
higher-redshift component (i.e., corresponding to the galaxies with
original redshift centered at $\sim$21,000 \kms) is approximately
twice as large as for the lower redshift component, 2200 and 1100
\kms, respectively. Furthermore, the two 
components show no spatial distinction from each other and are spread
throughout the entire observed region of the HRS.

To quantify the likelihood of bi-modality in the redshift
distribution,  we first assess the likelihood that a single Gaussian
provides an adequate fit to the data. At each PA, we determine the
residual redshift of all galaxies from the systematic
position-redshift trend, and then we employ KMM statistics to assess
the likelihood of a two-Gaussian versus a single-Gaussian fit
\citep{ash94}.  For {\it all} position angles, a common covariance,
two-Gaussian fit is preferred to a single Gaussian  with a high degree
of confidence ($>$ 99\%). The average peak-to-peak separation of the
two components over the entire PA range is 3014 $\pm$ 712 \kms, with
the separation along the best-fit line (PA$=-$80\arcdeg) being 3003
$\pm$ 174 \kms. Next, we utilize $\chi ^2$ statistics to test  
the goodness-of-fit for two Gaussian distributions with differing FWHM
as a function of PA. The reduced $\chi ^2$ values range from 
0.95 to 4.91 with the best fit value at PA$=-$60\arcdeg. In summary, the
statistical tests confirm the bi-modal nature of the HRS redshift
distribution, with the clearest distinction between the two redshift
components occurring along the principal spatial-redshift axis of the
supercluster. 

Finally, we have utilized the same KMM statistical methods to assess the
redshift distribution of clusters in the HRS with known redshift.  For the
clusters, the best fit correlation axis was found at
PA$=-$50\arcdeg. However when considering the broad nature of the
correlation$-$PA relationship for clusters, we used the best fit line
from the inter-cluster galaxies at a PA of $-$80\arcdeg\ (cf., Figure
\ref{f7}). We fitted the systematic position-redshift trend of
the galaxies and found the residual mean redshift for each cluster from
that trend.  The resulting histogram of cluster residual redshift was
plotted in Figure \ref{f9} (right).  We then applied KMM
statistics to the cluster histogram. Unlike the test on the galaxies,
the cluster histogram showed no clear signature of a bi-modal redshift
distribution.  Specifically, the cluster redshifts fitted a bi-modal
distribution with $\sim$75\% confidence as compared to a single
Gaussian distribution.  However, as can be seen in Figure
\ref{f9} (right), the cluster redshift data were sparse and
little could be concluded from their redshift distribution. 

\section{Comparisons with the Shapley Concentration}
      
The HRS is generally referred to as the second largest supercluster
within 200 Mpc, second only in mass to the Shapley Supercluster
(SSC) \citep{hud99}. Since the SSC is both well-studied and the
most comparable supercluster in the local universe, we use it
as a benchmark for assessing the properties of the HRS. The comparison
between these two largest structures is somewhat hindered by the fact
that most of the SSC studies combine inter-cluster and cluster
galaxies, while our 6dF data for the HRS samples only the inter-cluster
galaxies.

\subsection{Extent and Overdensity}
We begin by comparing the kinematic extent of the HRS (from 17,000 to
22,500 \kms) with that of the SSC. The velocity boundaries of
the entire SSC are generally cited as extending from 8,000 to 18,000 \kms\
\citep{qui00,dri04}. To put the SSC on the same quantitative
footing as the HRS, we compare the cluster populations of the two
superclusters.  Specifically, when compared with the 18 ACO clusters 
found in the HRS by \citet{zuc93}, the same authors find 24 ACO
clusters in the SSC, while \citet{ein01} find 25. Hence the numbers of
clusters in the SSC are comparable to, perhaps slightly larger than,
those in the HRS. For the 24 ACO clusters combined from these studies, we
used published mean redshift data from \citet{qui00} to calculate a
comparative kinematic extent for the SSC. We determine the
FWHM of the redshift distribution of the SSC clusters to be $\sim$6000
\kms, very similar to the $\sim$5500 \kms\ found for the HRS.  As is
discussed below, the redshift distribution of the SSC clusters is distinctly
bi-modal, thus the FWHM metric is rather an oversimplification of a
complex environment. However, the basic result is that the HRS and SSC
are similar in regard to their total number of ACO clusters and
overall kinematic extent.

Next, we seek to make a valid comparison between the inter-cluster
overdensities of the SSC and the HRS. Three studies have examined in
detail the inter-cluster overdensity of the SSC
\citep{dri99,bar00,dri04}. Of these, only \citet{dri04} considers a similar
area on the sky, so we draw a comparison with this study. The SSC
inter-cluster overdensity is 3.3 $\pm$ 0.1 over 151 deg$^2$, as
compared with an overdensity of 2.4 that we find for the HRS. A radius
of 0.5\degr\ ($\sim$ 2 Mpc at SSC mean redshift) was excised around
ACO clusters within the survey area. This radius in the SSC
corresponds to 0.35\degr\ at the HRS redshift. Consequently, the HRS
sample is slightly more restrictive in selecting only inter-cluster
galaxies, but this is likely a small difference.  Overall, we find a
somewhat smaller, but similar, overdensity in the HRS compared to that
in the SSC. 

In addition, we compare the total mass in the HRS inter-cluster galaxies
to that in the SSC.  Given the differences between the HRS and SSC
studies in overdensity (2.4/3.3), angular survey region (107/151
deg$^2$), and relative distance ($\sim$ 20,000/15,000 \kms), we
conclude that the total masses of the inter-cluster regions of the HRS
and SSC are virtually identical.  Thus our data indeed support the
conclusion of previous studies of the distribution 
of galaxy clusters \citep{zuc93,hud99,ein01} that the SSC and HRS 
constitute the two largest mass concentrations in the local universe.

\subsection{Morphological Considerations}
In \S3.2, we found an overall spatial-redshift trend in the HRS, in that 
a systematic increase in redshift is present with increasing position along
a SE$-$NW axis. Bardelli et al. (2000) have fitted a plane in ($\alpha$,
$\delta$, $cz$) space to their inter-cluster observations in the
SSC. They note a $\sim$3000 \kms\ increase in average galaxy velocity
along the best fit plane over the 8\arcdeg\ (40 Mpc) region, a result
reminiscent of the position-redshift tilt in the HRS. However, when
the area on the sky is expanded \citep[cf., Figure 4 of][]{bar00} to
include the inter-cluster galaxies in \citet{dri99}, the main peak of
the galaxy distribution shifts by 7 Mpc, and the entire
distribution is broadened.  In short, while there appears to be a
kinematic gradient in the SSC, it is not clear whether that feature
extends over the entire region of the supercluster.

As is discussed in \S3.3, when the spatial-redshift trend along the
PA=$-$80\arcdeg\ axis in the HRS is fitted and removed, the redshift
distribution is bi-modal (Figure \ref{f9}). In fact, the
bi-modal signature is observed even in the original redshift
histogram in Figure \ref{f4}. Redshift bi-modality is also
strikingly evident in the SSC \citep[cf., Figure 6 of][Figure
  5]{dri04,qui00}. There is a lower redshift component  to the SSC (at 
$\sim$8,000 $-$ 12,000 \kms) that is quite distinct from the higher
component at $\sim$14,000 $-$ 18,000 \kms. While it 
was originally thought that the SSC redshift components are
substantially different in size, an extensive follow-up study by
\citet{dri04} reveals the {\it inter-cluster} populations of the two
components to be roughly equal \citep[cf., Figure 5
of][]{dri04}. On the other hand, the
distribution of the {\it clusters} within the SSC is also bi-modal
and more heavily weighted to the higher redshift component at
13,000 $-$ 18,000 \kms.  Specifically, 16 clusters have redshifts above
13,500 \kms, while only 6 have redshifts below 12,500 \kms. The
higher redshift component coincides with what is designated by
\citet{rei00} as  the collapsing ``Central Region,'' centered on the
cluster A3558. As a result, the higher redshift component dominates
when both  the cluster and inter-cluster galaxies are considered. More
reliable redshift data for the clusters in the HRS is probably
required before a definitive statement can be made about their redshift
distribution, but the available data (cf., Figure \ref{f9},
Right panel) indicate that such a 3:1 imbalance in cluster numbers
between lower and higher redshift components is not present. 

Although our observations reveal a distinct arrangement of field
galaxies marking the HRS, the extent and/or boundaries of the
supercluster are not easily determined. In fact, percolation and
friends-of-friends algorithms include other clusters in the HRS
besides those listed in Table \ref{tb3} \citep{kal98,ein02}. Recent
studies of the SSC cover a similar area on the sky and also leave some
ambiguity as to the spatial and kinematic extent of the supercluster
\citep{qui00,dri04}. It is quite possible that the boundaries of the
HRS extend beyond the region surveyed by us with 6dF.  

\section{CONCLUSIONS}

We have obtained optical redshift data for 547 inter-cluster galaxies
in the region of the Horologium-Reticulum Supercluster (HRS). This
extensive coverage of the inter-cluster galaxies provides
an opportunity to define large-scale kinematic structures within the
HRS. Our initial result is the detection of a main concentration of
inter-cluster galaxies from 17,000 $-$ 22,500 \kms, which we refer to as
the HRS kinematic extent. This was followed by the comparison of our
observations with a smooth, homogeneous galaxy distribution. An
overdensity of 2.4 was calculated, or $\delta \rho / \bar{\rho}
\sim$ 1.4, which reveals that the HRS complex has entered the
non-linear regime. Through visual inspection of redshift slices, 
reinforced with correlation analysis, the galaxies within the
kinematic extent are found to exhibit a significant trend in redshift with
position along a SE$-$NW axis in the sense that redshift increases by
$\sim$1500 \kms\ along this axis. Furthermore, the resulting position angle
of the trend is closely aligned with that found in the clusters within
the HRS. In addition, when the kinematic trend found above is
accounted for and removed, we find a distinct bi-modality to the
redshift distribution of the inter-cluster galaxies within the
HRS. Thus, the HRS can be viewed as consisting of two major
components in redshift space, separated by 2500 \kms\ (35 Mpc), each
with a similar position-redshift tilt at the same position angle.     

We thank Fred Watson, Paul Cass, and the 6dFGS team for
supervising and conducting our observations, as well as Keith Ashman
for his quick response with the KMM code. We also thank Rien
van de Weygaert and Saleem Zaroubi at the Kapteyn Institute,
Groningen, NL for helpful discussions. We acknowledge useful comments
from the referee, which led to considerable improvements in the final
paper. MCF acknowledges the support of a NASA Space Grant Graduate
Fellowship at the University of North Carolina-Chapel Hill. RWH
acknowledges grant support from the Australian Research Council. A
portion of this work was supported  by NSF grants AST-9900720 and
AST-0406443 to the University of North Carolina-Chapel Hill.

\newpage
\onecolumn

\newpage
\begin{deluxetable}{crrccclcr}
\tabletypesize{\scriptsize}
\tablecaption{Spectroscopic Observations \label{tb1}}
\tablewidth{0pt}
\tablehead{
\colhead{Date} & \colhead{$\alpha_{2000}$}   & \colhead{$\delta_{2000}$} &
\colhead{ID} &
\colhead{Field No.}  & \colhead{Grating} & \colhead{t$_{exp}$(s)} &
\colhead{Seeing} & \colhead{S/N}
}
\startdata
 2002 Oct 31&02 55 57.9 &$-$50 18 20 &3110 & 198,199&580V &4$\times$1200&
2$-$3\arcsec & 7.5\\
& & & &154 &425R   &4$\times$600&3$-$5\arcsec & 9.0\\
2002 Nov 01&02 55 57.9 &$-$51 38 12 &0111 & 198,199 &580V  &
4$\times$1200&1$-$2\arcsec & 9.0\\
& & & &154 &425R  &4$\times$600&1$-$2\arcsec & 11.7\\
2002 Nov 03&03 02 00.4 &$-$46 18 17 &0411 &247, 248  &580V  &
4$\times$600&3$-$5\arcsec & 4.6\\
2002 Nov 04 &03 02 00.5 &$-$46 18 15 &0411 &247, 248  &425R
&4$\times$600&2$-$3\arcsec & 10.8\\
& & & & &580V   &4$\times$1200&3$-$4\arcsec & 8.7\\
2002 Nov 05&03 24 57.6 &$-$50 58 17 &0511 &200  &425R &
4$\times$600&2$-$3\arcsec & 12.6\\
& & & & &580V  &4$\times$1200 &2.2\arcsec & 9.8\\
2002 Nov 06&03 17 55.5 &$-$55 48 06 &0611 &155  &425R
&4$\times$600&2$-$3\arcsec& 10.8\\
& & & & &580V   &4$\times$1200 &3$-$4\arcsec & 7.8\\
2002 Nov 07&03 28 54.8 &$-$56 58 04 &0711 &155, 156  &425R  &4$\times$600
&1$-$2\arcsec & 10.5\\
& & & & &580V   &6$\times$1200 &3$-$4\arcsec , cloudy & 8.6\\
2002 Nov 08&03 33 02.2 &$-$46 28 04 &0811 &200, 248  &425R  &
4$\times$600&1$-$2\arcsec & 5.4\\
& & & &249 &580V  &4$\times$1200&1$-$2\arcsec & $-$ \\
 \enddata

\tablecomments{Numbers in parentheses refer to the column numbers.
  (1) Date of observation, (2) Right Ascension of the
  field center in hours, minutes, and seconds (J2000), (3) Declination of the
  field center in degrees, arcminutes, and arcseconds (J2000), (4)
  Identification number as found in Figure \ref{f1},
  (5) Schmidt field number, (6) Grating, (7) Exposure time, (8)
  Approximate seeing, (9) Average signal-to-noise.} 

\end{deluxetable}

\newpage
\begin{deluxetable}{ccccccc}
\tabletypesize{\scriptsize}
\tablecaption{ Velocity Data for 6dF Galaxy Spectra \label{shorttb2}}
\tablewidth{0pt}
\tablehead{
\colhead{ID} & \colhead{$\alpha_{2000}$}   & \colhead{$\delta_{2000}$} &
\colhead{b$_J$} & \colhead{Reference}  & \colhead{$cz$ (\kms)} &
\colhead{$\sigma_{cz}$} 
}
\startdata
HRS J024113$-$501154 & 02 41 13.42 & $-$50 11 54.1 & 13.41 & e & 27662 & 40 \\
HRS J024126$-$514012 & 02 41 26.02 & $-$51 40 12.0 & 17.36 & e & 30523 & 86 \\
HRS J024141$-$524151 & 02 41 41.38 & $-$52 41 51.7 & 17.17 & e & 14078 & 52 \\
HRS J024141$-$505106 & 02 41 41.86 & $-$50 51 06.1 & 16.93 & e & 18818 & 47 \\
HRS J024213$-$514333 & 02 42 13.22 & $-$51 43 33.2 & 17.45 & e & 22710 & 50 \\
... & ... & ... & ... & ... & ... & ... \\
 \enddata

\tablecomments{Numbers in parentheses refer to the column numbers.
  (1) IAU name, (2) Right ascension in hours, minutes, and seconds
  (J2000), (3) Declination in degrees, arcminutes, and arcseconds
  (J2000), (4) b$_J$ magnitude as listed in SuperCOSMOS, (5) ``a''=
  absorption lines used to calculate redshift, ``e''= emission lines
  used to calculate redshift, ``ae''= both absorption and emission
  lines used, (6) Velocity, $cz$, (7) Velocity error. The complete
  version of this table is in the electronic edition of the Journal.
  The printed edition contains only a sample. }
\end{deluxetable}

\newpage
\begin{deluxetable}{ccccccc}
\tabletypesize{\scriptsize}
\tablecaption{ Clusters of Known Redshift in the Observed Region
  \label{tb3}} 
\tablewidth{0pt}
\tablehead{
\colhead{Cluster} & \colhead{$\alpha_{2000}$}   & \colhead{$\delta_{2000}$} &
\colhead{List} &
\colhead{Redshift}  & \colhead{cz (\kms)} & \colhead{Source} 
}
\startdata
A3047 & 02 45.2 &$-$46 27.0	& & 0.0950 & 28500 &1\\
A3074 & 02 57.9 & $-$52 43.0	& B & 0.0730 & 21900 & 1\\
A3078 & 03 00.5 & $-$51 50.0 & B & 0.0648 & 19440 & 2\\
A3093 &	03 10.9	& $-$47 23.0 & B & 0.0830 & 24900 & 2\\
M031027&  03 11.9 & $-$52 54.0&& 0.0570& 17088 & 1\\
A3100 &	03 13.8	& $-$47 47.0 & B & 0.0629&	18870  &2\\
A3104 &	03 14.3 & $-$45 24.0 & B & 0.0730	&21900	 & 1\\
A3106 & 03 14.5 & $-$58 05.0 & B & 0.0639 & 19170 & 2\\
A3108 & 03 15.2 & $-$47 37.0 & B & 0.0625 & 18750 & 3\\
A3109 & 03 16.7 & $-$43 51.0 & B & 0.0920 & 27580 & 3\\
A3110 &	03 16.5	& $-$50 54.0 & B & 0.0749 & 22470 & 2\\
A3111 & 03 17.8 & $-$45 44.0	& E & 0.0775	& 23250 & 4\\	
A3112 & 03 17.9 & $-$44 14.0	& B & 0.0750 & 22500 & 4\\
S0339& 03 19.0 & $-$53 57.4& & 0.0546& 16369 & 1\\
S0345 & 03 21.8 & $-$45 32.3 && 0.0700 & 20985& 1\\
A3120 & 03 21.9 & $-$51 19.0	& B & 0.0690 & 20700 & 2\\
M391 & 03 22.3 & $-$53 11.3& &0.0780& 23384 & 1\\
A3123 & 03 23.0 & $-$52 01.0	& B & 0.0644 & 19320 & 2\\
M03233&  03 24.8 & $-$58 35.1& &0.0670&  20086 & 1\\
A3125 & 03 27.4 & $-$53 30.0	& B & 0.0589 & 17670 & 5\\
M399 &  03 28.4& $-$53 01.3&& 0.0600& 17988 &1\\
A3126 & 03 28.7 & $-$55 42.0 & & 0.0856 & 25680 & 4\\
S0356& 03 29.6& $-$45 58.8&& 0.0720& 21585& 1\\
A3128 &	03 30.2	& $-$52 33.0 & B & 0.0599 &	17970 & 4\\
A3133 &	03 32.7 & $-$45 56.0 & B & 0.0543	& 16290 & 2\\
M421&   03 35.5& $-$53 40.9& &0.0630& 18887 &1\\ 
A3144 & 03 37.1 & $-$55 01 0	& & 0.0443 & 13290 & 2\\
M433 &  03 41.1& $-$45 41.5&&  0.0660& 19786 &1\\ 
A3158 &	03 43.0 & $-$53 38.0 & B & 0.0597 & 17910 & 4\\
A3164 &	03 45.8 & $-$57 02.0 & B & 0.0570 & 17100 & 2\\
\enddata

\tablecomments{Numbers in parentheses apply to column numbers. (1)
  A-ACO, S-poor clusters from ACO, M- APM Galaxy Survey; (2) Right
  Ascension in hours and minutes (J2000); (3) Declination in degrees and 
  minutes (J2000); (4) B-HRS membership listed in both Zucca et
  al. (1993) and Einasto et al. (2001), E-HRS membership listed only
  in Einasto et al. (2001); (7) $^1$\citet{dal94},
  $^2$\citet{str99}, $^3$\citet{den95}, $^4$\citet{kat98},
  $^5$\citet{cal97}. }  
\end{deluxetable}

\end{document}